\documentclass[sigconf]{acmart}

\usepackage{caption}
\usepackage{subcaption}

\AtBeginDocument{%
  \providecommand\BibTeX{{%
    \normalfont B\kern-0.5em{\scshape i\kern-0.25em b}\kern-0.8em\TeX}}}

\setcopyright{acmcopyright}
\copyrightyear{2020}
\acmYear{2020}
\acmDOI{}

\acmConference[Workshop on Mental Models of Robots, HRI '20]{Workshop on Mental Models of Robots at HRI 2020}{March 23, 2020}{Cambridge, UK}
\acmBooktitle{Workshop on Mental Models of Robots at HRI 2020,
  March 23, 2020, Cambridge, UK}
\acmPrice{}
\acmISBN{} 

\begin{document}

\title{Natural Language Interaction to Facilitate Mental Models of Remote Robots}

\author{Francisco J. Chiyah Garcia}
\author{Jos\'e Lopes}
\author{Helen Hastie}
\email{{fjc3, jd.lopes, h.hastie}@hw.ac.uk}
\affiliation{%
  \institution{School of Mathematical and Computer Sciences, Heriot-Watt University}
  \city{Edinburgh}
  \state{United Kingdom}
  \postcode{EH14 4AS}
}

\renewcommand{\shortauthors}{Chiyah Garcia et al.}

\begin{abstract}

Increasingly complex and autonomous robots are being deployed in real-world environments with far-reaching consequences. High-stakes scenarios, such as emergency response or offshore energy platform and nuclear inspections, require robot operators to have clear mental models of what the robots can and can't do. 
However, operators are often not the original designers of the robots and thus, they do not necessarily have such clear mental models, especially if they are novice users. 
This lack of mental model clarity can slow adoption and can negatively impact human-machine teaming.  
We propose that interaction with a conversational assistant, who acts as a mediator, 
can help the user with understanding the functionality of remote robots and increase transparency through natural language explanations, as well as facilitate the evaluation of operators' mental models. 

\end{abstract}




\maketitle

\section{Introduction}

\begin{figure}
\centering
  \includegraphics[width=.65\linewidth]{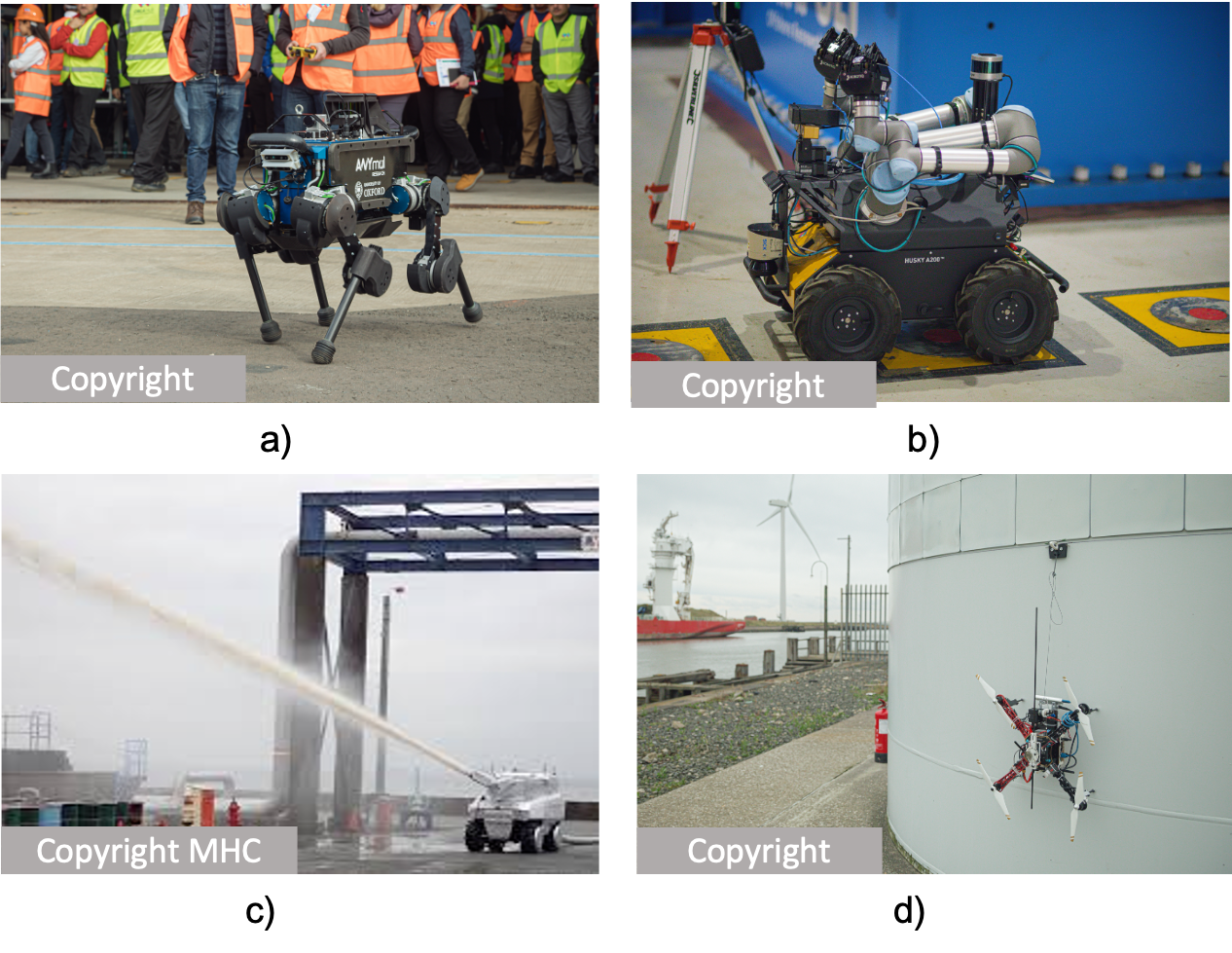}

\caption{Figure showing examples robots used for remote operation.} 
\label{fig:robots}
\vspace*{-7mm}
\end{figure}

Robots and autonomous systems are being deployed in remote and dangerous environments, such as in nuclear plants or on offshore energy platforms
for inspection and maintenance.  These robots are important as they keep operators out of harm's way  \cite{Hastie2018,Jinke2017,Kwon2012,Nagatani2018,Shukla2016,Wong2017}. However, to date there exists no single robot that has all the functionality to perform the variety of tasks required in these domains. For example in an offshore emergency response scenario, robots need to firstly inspect the emergency area (e.g. with a ground robot with a camera and other sensors); secondly, resolve the emergency (e.g. with a heavy ground robot that can put out a fire); and finally, inspect the damaged area (e.g. with a drone collecting aerial images). See Figure \ref{fig:robots} for examples of such robots (where images a, b and d are images of robots from the ORCA Hub \cite{Hastie2018}).  Until a single robot can do all these tasks, the operator will be required to manage multiple robots, all functioning differently and performing tasks in different ways. This problem is confounded by the advent of robots that can adapt, with their functionality and behaviour changing continuously \cite{Cully2015}. Furthermore, remotely-controlled robots often instil less trust than those co-located \cite{Bainbridge2008,Li2015}, thus it is essential that operators maintain an appropriate mental model of the robot. This is a huge burden on the operator to gain and maintain such clear mental models of each of these robots and to task and manage them effectively.  This means that only highly skilled operators, using a variety of interfaces,  would be able to control the robots and this could potentially hinder general adoption. 


Ideally, the operator would interact with the remote robots at a task level (e.g. ``go and inspect this area and resolve any emergency that you find''). To be able to do this would involve multiples stages including: understanding the user's intent, communicating the intent of the multiple robots in a way that is easy for the operator to understand and maintaining situation awareness, including a level of transparency.    We, therefore, propose a conversational assistant, called MIRIAM, who is able to act as an intermediary. As dialogue and natural language is universal, it reduces the need for specific robot training and facilitates the formation of high-fidelity mental models through natural language explanations and interaction.




\section{Remote Robots and Mental Models}

A clear understanding of the actions and reasoning of a robot is crucial for the operator and increases the robot's transparency, an important factor in explainability \cite{wortham2017}, preventing issues such as wrong assumptions, misuse or over-trust. It also helps the operator build a more faithful mental model of the robot, which comes with increased confidence and performance \cite{lebras18,Lim2009}. In cognitive theory, mental models provide one view to explain how a person thinks and reasons, either functionally (e.g. understanding what a robot does) or structurally (e.g. how a robot works) \cite{JohnsonLaird80}, 
and strongly impact how a person perceives and uses a system.
Therefore, it is essential to build and maintain a high-fidelity mental model, particularly for robots that are remote where face-to-face
interaction is not physically possible and 
with robots that are involved in high stakes and rapidly evolving scenarios, such as emergency response.


Although the designers of the system have a more complete and faithful mental model, it is difficult to transmit this so that users can fully understand the system. \cite{norman1988psychology} suggests that this discrepancy between the designer and the user's mental models appears because the designer does not talk directly to the user. Instead, the designer conveys its mental model through the system itself, which is a materialisation and thus \textit{open to interpretation}. This transmission of information to the user is a challenge for remote robots, which are typically represented by a moving dot on a screen and therefore are unable to implicitly communicate capabilities that in-proximity social robots might be able to. 

Previous work has looked into doing this through communicating the robot's actions and reasoning through natural language. Recently, intelligent assistants have been explored as a way to monitor robots and give operators information that is easy to digest \cite{RobotVerbalization, LemonWITAS, Misu2013}. Furthermore, fully interactive intelligent assistants, such as Amazon Alexa and Google home,  aim to imitate the naturalness of human-to-human conversations by enabling interaction through natural language, including informal chitchat. As these assistants are becoming more commonplace, we believe that they can be used to facilitate a wider range of interactions, including those in the workplace and operators of machinery and robots, for example, in factories and for remote scenarios, as discussed here.

\vspace*{-4mm}

\section{The MIRIAM System}

The MIRIAM intelligent assistant \cite{HastieICMI17} is a typed chat or spoken dialogue system that uses natural language to interact with several remote autonomous vehicles, including drones, ground and underwater vehicles. It provides operators with status updates, alerts and explanations of events in mixed-initiative conversation and allows them to process the operator's queries and act on them. The system has been used to help with operator training and to investigate the effects of trust and situation awareness of operators with autonomous vehicles in the offshore domain \cite{RobbICMI18}. As reported in \cite{ChiyahINLG18}, explanations provided by the assistant marginally improved the mental model of operators in terms of what the autonomous vehicles were doing (functionally) and how the autonomous vehicles worked (structurally). There were also significant differences in how mental models evolved over time during the study depending on what was communicated and how, suggesting that the amount of content and how it is conveyed to the operator is very important. We have also shown, in previous work, that intelligent assistants can help to maintain situation awareness of robot activities through communication of updates and alerts \cite{RobbICMI18}.

Our system directly interacts with the autonomous robots illustrated in Figure \ref{fig:robots}a,b,d, and controls them, obtaining updates from them. Further processing of these updates enables the MIRIAM system to produce explanations that are then communicated to the operator (e.g. a robot with low battery or without a camera cannot be sent to inspect an area), thus increasing the transparency between the robots and the operator. It maintains a dynamic world view and is thus able to constrain the interaction and advise the operator on which robots to use based on their capabilities, standard operating procedures and current world view. This process in itself helps to develop and maintain the operator's mental model.




Estimating one's mental model and evaluating any increases in clarity is a very challenging task, even between humans. Our previous work with MIRIAM showed how we can estimate the mental model of the participants by asking them to rate statements that measured several dimensions about their understanding of what was happening and why \cite{ChiyahINLG18}. These measurements were taken whilst the autonomous vehicles performed tasks, both before and after the intelligent assistant had provided information and an explanation.
\cite{norman1983a} argued that mental models can be incomplete, unstable, contradictory and change over time, amongst other properties. This can make mental models inconsistent and, according to \cite{Rouse1986}, particularly difficult to track in experiments. We have showed, in prior work, however, that this was possible through a constrained task.



\begin{figure}
\centering
  \includegraphics[width=0.4\linewidth]{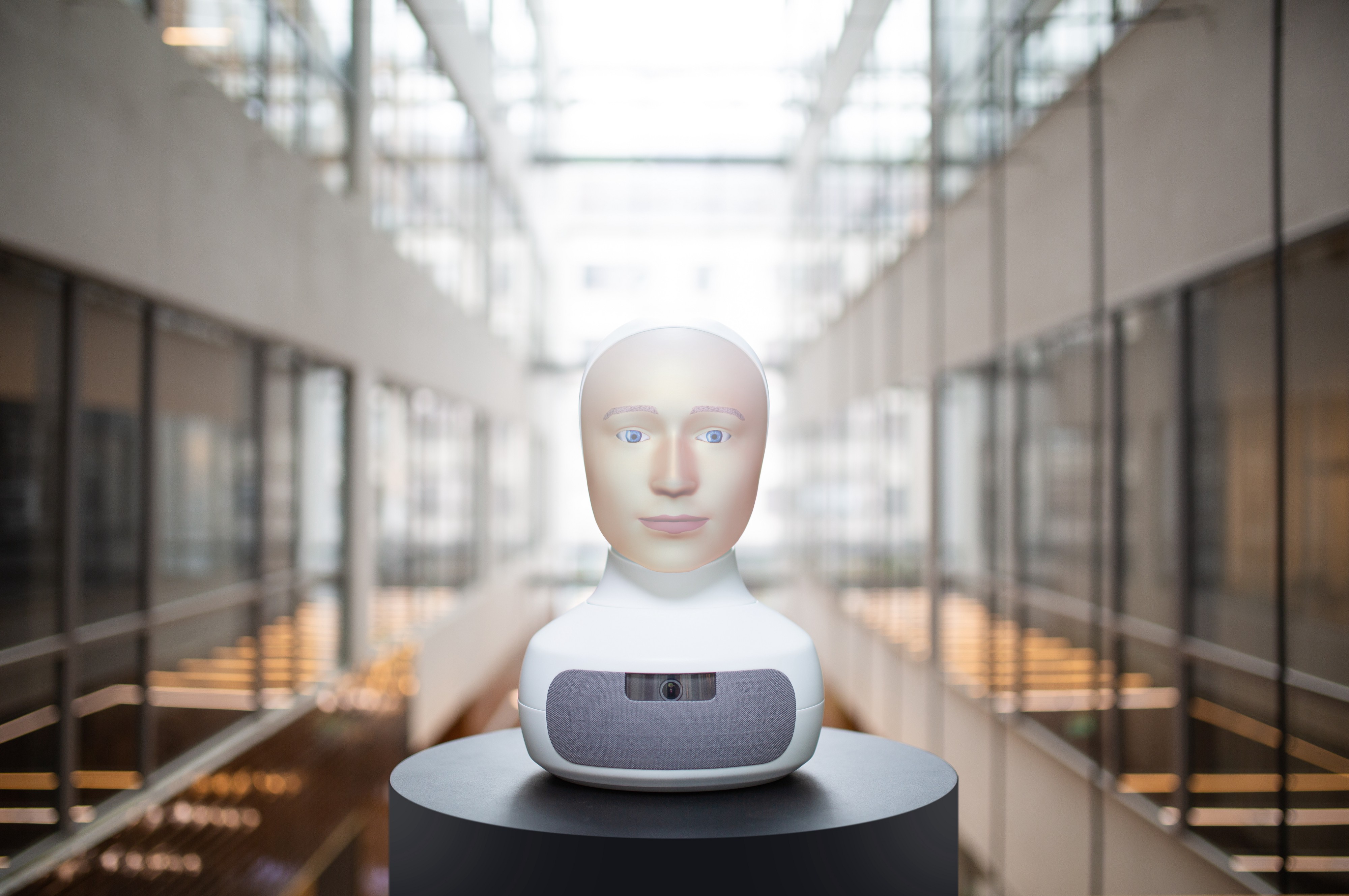}

\caption{Furhat social robot used to embody the MIRIAM conversational agent.}
\label{fig:furhat}
\vspace*{-7mm}
\end{figure}



\vspace*{-1mm}

\section{Future work and Conclusion}

The future of human-machine cooperation lies in effective interaction between humans and robots. Relating to the Theory of Mind \cite{Baron-Cohen1988, Premack1978}, a conversational agent that can understand what the user knows or does not know and corrects wrong assumptions could assist further in such high-stake scenarios. As discussed above, initial work has shown that one can evaluate techniques designed to improve mental models. However, estimating the user's general knowledge and expertise level related to robots and a scenario and task is an area for future research. 

Communicating what the robot/system can do explicitly is not always an effective method, as it may repeat what the user already knows unnecessarily. One method of subtly doing this is through social cues. Initial work \cite{Lopes2019} has looked at this by embodying the conversational assistant into the form of a Furhat social robot (Figure \ref{fig:furhat}). 
This enables the use of social cues that extend existing dialogue and pragmatic cues
in the spoken dialogue system to include 
visual social cues, such as shared gaze and facial expressions, or auditory cues such as prosody. 
Rather than continuously explicitly stating the functions of the robot, we are investigating if such social cues would lead to more natural and time-efficient interaction and would reflect more the way humans interact with each other.  
We are also currently exploring if such social robots can facilitate adoption by increasing trust and understanding of the user through such socially enhanced interaction.

\begin{acks}

This work was supported by the EPSRC funded ORCA Hub \\(EP/R026173/1, 2017-2021). We thank the anonymous reviewers for their insightful comments.

\end{acks}

\bibliographystyle{ACM-Reference-Format}
\bibliography{main}


\end{document}